----------------------------- Start of body part 2

\documentstyle{article}

\newcommand{\be}{\begin{equation}}
\newcommand{\ee}{\end{equation}}
\newcommand{\bea}{\begin{eqnarray}}
\newcommand{\eea}{\end{eqnarray}}
\newcommand{\uq}{\mbox{${\cal U}_q(sl(2))$}}
\newcommand{\uqa}{\mbox{${\cal U}_q(\widehat{sl(2)})$}}

\title{On Integrable Quantum Group Invariant
Antiferromagnets\thanks{Lecture given by C.G. at the Winter School on
Theoretical Physics, Wroclaw, Poland, February 1992.
Keywords: Integrable models. Open chains. Quantum groups.
1991 MSC: 82B23, 17B37.
PACS: 05.50.+q, 02.90.+p, 75.10 Jm}}

\author{R. Cuerno\thanks{e--mail: IMTRC59@CC.CSIC.ES. Supported by a PFPI
grant of the spanish MEC.}, G. Sierra\thanks{e--mail: SIERRA@CC.CSIC.ES} \\
{\em Instituto de Matem\'aticas y F\'{\i}sica Fundamental, CSIC} \\
{\em Serrano 123, E-28006 Madrid, SPAIN} \\
C. G\'omez\thanks{e--mail: GOMEZC@SC2A.UNIGE.CH. Permanent
address: IMFF, CSIC.} \\
{\em D\'epartment de Physique Th\'eorique} \\
{\em Universit\'e de Gen\'eve, CH--1211 Geneva, SWITZERLAND}}

\date{}

\begin{document}
\maketitle

\begin{abstract}
A new open spin chain hamiltonian is introduced. It is both
integrable (Sklyanin's type $K$ matrices are used to achieve this)
and invariant under ${\cal U}_{\epsilon}(sl(2))$ transformations
in nilpotent irreps for $\epsilon^3=1$. Some considerations on
the centralizer of nilpotent representations and its representation
theory are also presented.
\end{abstract}

\vskip-16.0cm
\rightline{{\bf IFF-5/92}}
\rightline{{\bf May 1992}}
\vskip2cm

\newpage
The most direct way to get into the physics associated with quantum groups is
certainly the study of quantum mechanical systems possesing a quantum
group symmetry. Some examples of this kind are already known in the
context of one dimensional spin chains \cite{PS}. The simplest one is the
XXZ spin--1/2 chain with boundary conditions:
\begin{equation}
H = \sum_{i=1}^{N-1} \sigma_i^x \sigma_{i+1}^x + \sigma_i^y \sigma_{i+1}^y
   + \frac{q + q^{-1}}{2} \sigma^z_i \sigma_{i+1}^z + \frac{q - q^{-1}}{2}
   (\sigma_1^z - \sigma_N^z)
\label{1}
\end{equation}

\noindent
which is invariant under ${\cal U}_q(sl(2))$ transformations. The integrable
version of the spin one Heisenberg model with non vanishing anisotropy is the
Zamolodchikov--Fateev hamiltonian \cite{ZF}:
\begin{eqnarray}
H^{ZF} & = & \sum_{i=1}^{N-1} H_{i,i+1}^{ZF} \nonumber \\
& = & \sum_{i=1}^{N-1} S_i^x S_{i+1}^x + S_i^y S_{i+1}^y + \frac{q^2 +
q^{-2}}{2} S_i^z S_{i+1}^z - (S_i^x S_{i+1}^x + S_i^y
S_{i+1}^y)^2 \nonumber \\
& & -\frac{q^2 + q^{-2}}{2}(S_i^z S_{i+1}^z)^2
+ (1-q-q^{-1}) [(S_i^x S_{i+1}^x +
S_i^y S_{i+1}^y) S_i^z S_{i+1}^z + \leftrightarrow] \nonumber \\
& & + (\frac{q^2 + q^{-2}}{2}
-1) ((S_i^z)^2 + (S_{i+1}^z)^2)
\label{2}
\end{eqnarray}

\noindent
The hamiltonian $H_{i,i+1}^{ZF}$ is given by the logarithmic derivative of the
spin one $R$--matrix $R^{(s=1)}(u)$ of the affine Hopf algebra
\uqa\ . In order to make the ZF hamiltonian invariant
under global \uq\ transformations the following boundary term should be added
\be
H^{(B)} = \frac{q^2 + q^{-2}}{2} (S_N^z - S_1^z)
\label{3}
\ee

\noindent
Integrability in a box requires that the Sklyanin reflection
operators $K_{\pm}(u)$, which describe the scattering with the
wall obey the relations \cite{S}:
\be
R_{12}(u-v) \stackrel{1}{K}_-(u)R_{12}(u+v) \stackrel{2}{K}_-(v)
= \stackrel{2}{K}_-(u) R_{12}(u+v) \stackrel{1}{K}_-(u) R_{12}(u-v)
\label{4}
\ee

\noindent
For the quantum group invariant hamiltonian $H^{ZF} + H^B$ it is
not hard to prove \cite{NMR} that:
\be
[H^{ZF} + H^B, t(u)] = 0
\label{5}
\ee

\noindent
where the box transfer matrix $t(u)$ is defined by:
\be
t(u) = {\rm Tr} (K_+(u) T(u) K_-(u) T^{-1}(-u))
\label{6}
\ee

\noindent
with $K_-$ satisfying (\ref{4}) for $R$ the spin one $R$--matrix
of \uqa\ and $K_+(u) = K_-(-u - \eta)$, $q = \exp \eta$. From
(\ref{5}) it follows the integrability of the
hamiltonian $H^{ZF} + H^B$. The monodromy matrix $T(u)$ in
(\ref{6}) is the one defined by the $s=1$ quantum $R$ matrix of \uqa.

In reference \cite{BGS} a one parameter family of
integrable deformations of (\ref{2}) was defined for $q = \epsilon$,
$\epsilon^3=1$ :
\bea
 H(\lambda) & = & 2 (\epsilon - \epsilon^{-1}) \sum_{i=1}^N
\frac{\lambda \epsilon + \lambda^{-1} \epsilon^{-1}}{2}
(S_i^x S_{i+1}^x + S_i^y S_{i+1}^y) \nonumber \\
& & - \frac{1}{2} S_i^z S_{i+1}^z -
(S_i^x S_{i+1}^x + S_i^y S_{i+1}^y)^2 +
\frac{1}{2}(S_i^z S_{i+1}^z)^2 \label{7} \\
& & + (\frac{\lambda \epsilon + \lambda^{-1} \epsilon^{-1}}{2} +
\omega) [(S_i^x S_{i+1}^x +
S_i^y S_{i+1}^y) S_i^z S_{i+1}^z + \leftrightarrow] - \frac{3}{2}
((S_i^z)^2 + (S_{i+1}^z)^2) \nonumber \\
& & - \frac{\lambda \epsilon - \lambda^{-1}
\epsilon^{-1}}{2(\epsilon - \epsilon^{-1})} (S_i^x S_{i+1}^x +
S_i^y S_{i+1}^y)(S_i^z + S_{i+1}^z) + \frac{\lambda^2
\epsilon^{-1} - \lambda^{-2} \epsilon}{2(\epsilon -
\epsilon^{-1})} (S_i^z + S_{i+1}^z) \nonumber \\
& \equiv & \sum_{i=1}^N H(\lambda)_{i,i+1} \nonumber
\eea

$$
\omega(\lambda) = \sqrt{(\lambda - \lambda^{-1})(\lambda
\epsilon^{-1} - \lambda^{-1} \epsilon)}
$$

\noindent
The hamiltonian $H(\lambda)_{i,i+1}$ is (up to a constant
factor) the logarithmic derivative of the quantum $R$--matrix
$R^{(\lambda)}(u)$ which intertwines nilpotent representations
of \uqa\ for $q=\epsilon$ \cite{BGS,BGSb}. Let us recall that
nilpotent representations of ${\cal
U}_{\epsilon}(\widehat{sl(2)})$ ($\epsilon^3 = 1$) are
three--dimensional irreducible representations transforming
under the central Hopf subalgebra as: $E^3 = F^3 = 0$, $K^3 =
\lambda^3$, with $\lambda$ a generic complex number. For
$\lambda = \epsilon^2$, which corresponds to the spin one
representation, $H(\lambda)$ coincides with
$2(\epsilon-\epsilon^{-1})H^{ZF}$ for $q=\epsilon$.

The problem we want to address in this lecture is the
definition of an integrable and quantum group invariant version
of (\ref{7}). Quantum group invariance is easily obtained adding
to (\ref{7}) the boundary term:
\be
H^B(\lambda) = \omega^2 (S_1^z - S_N^z)
\label{8}
\ee

\noindent
The hamiltonian $H(\lambda) + H^B(\lambda)$ coincides, for
$\lambda = \epsilon^2$, with $2(\epsilon-\epsilon^{-1}) (H^{ZF}+
H^B)$, which is already a good indication concerning
integrability. However, to attain it, we need to check for
$H(\lambda) + H^B(\lambda)$ the equivalent to equation (\ref{5})
with $K$ now being a solution to (\ref{4}) for $R$ the quantum
nilpotent $R^{\lambda}(u)$--matrix. The $K$--matrices for the
nilpotent $R^{\lambda}(u)$--matrix are:
\bea
K_-(u) & = & \frac{1}{\sinh \alpha_- \sinh(\alpha_- - \eta)} \nonumber \\
& \times & {\rm diag}(\sinh(u+\alpha_-) \sinh(u+\alpha_- - \eta),
-\sinh(u-\alpha_-) \sinh(u+\alpha_- - \eta), \nonumber \\
& & \sinh(u-\alpha_-) \sinh(u-\alpha_- + \eta)) \nonumber \\
K_+(u) & = & {\rm diag}(\sinh(u+ \eta -\alpha_+) \sinh(u-\alpha_+ - \eta),
\label{9} \\
& & -\sinh(u+\eta+\alpha_+) \sinh(u-\alpha_+ - \eta),
\sinh(u+\alpha_+) \sinh(u+\alpha_+ + \eta)) \nonumber
\eea

\noindent
with $\alpha_{\pm}$ free parameters and $\eta = 2 \pi {\rm
i}/3$. Note that these matrices possess precisely the same form
as those used in reference \cite{NMR} to prove the
integrability of the Zamolodchikov--Fateev spin one chain with
boundary terms. Using these $K$ matrices we derive for
$H(\lambda) + H^B(\lambda)$ the integrability condition
(\ref{5}) by showing that $H(\lambda) + H^B(\lambda)$ is proportional
to the second derivative, at the point $u=0$, of
the box transfer matrix $t(u)$, for $\alpha_{\pm} = \infty$.
Notice the diference with the ZF case where the hamiltonian is
given by the first logarithmic derivative of $t(u)$. The reasons
are that in our case, as it can be seen from (\ref{9}), $t(0) =
{\rm Tr} K_+(0)$ becomes zero, and that ${\rm
Tr}(\stackrel{0}{K}_+(0) H(\lambda)_{N0}) \propto
\stackrel{N}{{\bf 1}}$.

The nice thing about a quantum group invariant hamiltonian is
that most of the properties of the spectrum can be directly
derived from representation theory. So for instance, for the
hamiltonian (\ref{1}) we know that each energy eigenvalue is
associated with a given spin--j irrep of \uq\ and that it would
be (2j+1) times degenerate. The different irreps that can appear
in the spectrum are the ones obtained by decomposing
$\stackrel{N}{\bigotimes} V^{1/2}$. Moreover the different hwv's
transforming in the same representation j will define
irreducible representations of the centralizer of \uq, which for
spin 1/2 is given by the Hecke algebra. In the massless phase
($|q|=1$), the previous results provide, together with the
systematic use of the finite size technology \cite{C}, the basis
for the quantum group interpretation of conformal field theories
\cite{BPZ}. A similar study can now be done for the hamiltonian
$H(\lambda)+ H^B(\lambda)$ with the new features being
associated to the peculiarities of the representation theory at
roots of unit.

In what follows we will concentrate our analysis on the
structure of the centralizer for nilpotent representations of
${\cal U}_{\epsilon}(sl(2))$. Given a nilpotent representation
$V^{\lambda}$ we define the centralizer
$C^{\lambda}_N(\epsilon)$ as the algebra of endomorphisms $g:
\stackrel{N}{\bigotimes} V^{\lambda} \rightarrow
\stackrel{N}{\bigotimes} V^{\lambda}$
commuting with the quantum group action. To get the generators
of $C^{\lambda}_N(\epsilon)$ we first define the ``braiding
limit" of the quantum $R$--matrix $R^{\lambda \lambda}(u)$ of
the affine Hopf algebra ${\cal U}_{\epsilon}(\widehat{sl(2)})$
as follows:
\be
R^{\lambda}_{\pm} = \lim_{u \rightarrow \pm \infty} R^{(\lambda,
\lambda)}(u)^{r'_1 r'_2}_{r_1 r_2} \; e^{u(r_1-r'_2)}
\label{10}
\ee

\noindent
Elements in $C^{\lambda}_N(\epsilon)$ are then generated by
\be g_i^{\pm} = {\bf 1} \otimes \cdots \otimes
(R^{\lambda}_{\pm})_{i,i+1} \otimes \cdots \otimes {\bf 1}
\;\;\; i=1,\ldots,N-1
\label{11}
\ee

\noindent
Based on the spectral decomposition of $R_{\pm}^{\lambda}$ we
will assume that the set of generators $g_i$ is complete. In
order to get some insight into the structure of the centralizer we
will first consider the case $\epsilon^4=1$. In this case the
nilpotent representations are two--dimensional and the
``braiding limit" $R$ matrix is given by:
\bea
R^{\lambda} & = & \left(
\begin{array}{cccc}
\lambda & & & \\
& \lambda-\lambda^{-1} & 1 & \\
& 1 & 0 & \\
& & & -\lambda^{-1} \\
\end{array}
\right) \label{12} \\
 & = & \sigma^+ \otimes \sigma^- + \sigma^- \otimes \sigma^+ +
\frac{1}{2} \lambda^{-1} (\sigma^z \otimes {\bf 1}) + \frac{1}{2}
\lambda ({\bf 1} \otimes \sigma^z) +
\frac{\lambda-\lambda^{-1}}{2} {\bf 1} \otimes {\bf 1} \nonumber
\eea

\noindent
This $R$ matrix has two eigenvalues, $\lambda$ and
$-\lambda^{-1}$. The generators $g_i$ satisfy the Hecke relation:
\be
g_i^2 = (\lambda - \lambda^{-1}) g_i + {\bf 1}
\label{13}
\ee

\noindent
This means that the centralizer $C^{\lambda}_N(\epsilon)$ in the
case where $\epsilon= e^{\pi {\rm i}/2}$ gives us a
representation of the Hecke algebra $H_N(\lambda^2)$. It is well
known that for generic $q$ the irreducible representations of
$H_N(q)$  are in one to one correspondence with irreps of $S_N$,
see figure 1.
\begin{figure}
\vspace*{5cm}
\label{fig1}
\caption{}
\end{figure}
So we may ask which representations we get from the centralizer
$C^{\lambda}_N(\epsilon)$. At this point it is worthwhile to
recall that the centralizer $C^{1/2}_N(q)$ for the spin 1/2
representation of \uq\ is the quotient of a Hecke algebra
$H_N(q)$ by the relation
\be
g_i g_{i+1} g_i + g_i g_{i+1} + g_{i+1} g_i + g_i + g_{i+1} + 1
= 0
\label{14}
\ee

\noindent
which in turn is equivalent to reducing the allowed Young
tableaux to those with at most two rows.

In this case the $R^{1/2}$ matrix which intertwines two spin 1/2
irreducible representations of \uq\ is given by:
\be
R^{1/2} = \left(
\begin{array}{cccc}
q & & & \\
& 0 & q^{1/2} & \\
& q^{1/2} & q-1 & \\
& & & q \\
\end{array} \right)
\label{15}
\ee

\noindent
This $R$--matrix has also two eigenvalues, $-1$ and $q$, but the
main difference with respect to the $R$--matrix (\ref{12}) is
that in this case the multiplicities of the eigenvalues are 2
and 2, while for (\ref{12}) they were 1 and 3. The latter fact
can be understood from the decomposition rule $\frac{1}{2}
\otimes \frac{1}{2} = 0 \oplus 1$ (irrep 0 has dimension 1 and
irrep 1 is three--dimensional). More generally we see that
condition (\ref{14}) imposes a one to one relation between the
irreps of the centralizer $C^{1/2}_N(q)$ and the decomposition
into irreps of \uq\ of $\stackrel{N}{\bigotimes} V^{1/2}$. All
this means that the Brauer--Weyl theory also applies to the spin
1/2 representation of \uq.

For the centralizer $C^{\lambda}_N(\epsilon)$ we now try to
follow the same steps, namely to see which are the allowed Young
diagrams in figure 1 according to the decomposition
rules of nilpotent irreps. It was shown in \cite{GSR,A}  that the
decomposition rules of nilpotent irreps for generic values of
$\lambda$ are given by:
\be
\lambda \otimes \lambda = \bigoplus_{i=0}^{N'-1} \lambda^2 \epsilon^{-2i}
\label{16}
\ee

\noindent
where $\epsilon^N = 1$ ($N'=N$ for $N$ odd and $N'=N/2$ for $N$
even). In the case of $\epsilon=e^{{\rm i} \pi/2}$ equation
(\ref{16}) explains the multiplicities (2,2) of the eigenvalues
of the $R^{\lambda}$ matrix (\ref{12}), since $\lambda \otimes
\lambda = \lambda^2 \oplus (-\lambda^2)$ and both $\lambda^2$
and $-\lambda^2$ have dimension 2. Moreover the generators $g_i$
constructed out from $R^{\lambda}$ satisfy instead of the
relation (\ref{14}) the following one:
\bea
e_i^-e_{i+2}^-e_{i+1}^+e_i^+e_{i+2}^+ & = & e_i^-e_{i+2}^-e_{i+1}^-e_i^+
e_{i+2}^+ =0 \label{17} \\
e_i^{\pm} & \equiv & \frac{{\bf 1} \pm \lambda^{\pm 1}
R^{\lambda}}{1+\lambda^{\pm 2}} \nonumber
\eea

\noindent
which implies that the allowed Young diagrams, in the nilpotent
case, are those of ``corner" type:
\be
\vspace*{3cm}
\label{18}
\ee

\noindent
The Bratelli diagram describing the centralizer
$C_N^{\lambda}(\epsilon = e^{{\rm i} \pi/2})$ is that in figure 2.
\begin{figure}
\vspace*{4cm}
\label{fig2}
\caption{}
\end{figure}
We notice that the Young diagrams of the type (\ref{18}) are
precisely the only ones that contribute to the Alexander--Conway
polynomial as shown by Jones in \cite{J}. We would like also to
mention that the $R$--matrix (\ref{12}) coincides with the
intertwiner $R$ matrix for the fundamental representation of
${\cal U}_q(sl(1,1))$ (with $q$ replaced by $\lambda$), which
was used in reference \cite{KS} in order to construct the
Alexander polynomial. It has also been found in \cite{Ge} in the
context of boson representations of \uq\ . All this seems to
indicate alternative descriptions of the nilpotent irreps of ${\cal
U}_{\epsilon}(sl(2))$ for $\epsilon=e^{{\rm i} \pi/2}$.

Coming back to our problem, we can now compare the Bratelli in
figure 2 with the one we derive from the decomposition
rule (\ref{16}) in the case of $\epsilon=e^{{\rm i}\pi/2}$,
shown in figure 3.
\begin{figure}
\vspace*{3cm}
\label{fig3}
\caption{}
\end{figure}
It is then clear that the diagrams in figures 2 and
3 can be related under some identifications, as in
figure 4.
\begin{figure}
\vspace*{5cm}
\label{fig4}
\caption{}
\end{figure}
We now face two posibilities, either

\begin{itemize}
\item the set of generators $g_i$ given by (\ref{11}) is not
complet in the sense that the centralizer
$C^{\lambda}_N(\epsilon)$ is bigger, or

\item the centralizer $C^{\lambda}_N(\epsilon)$  is nothing but
the one generated by the $g_i$'s with Bratelli given by that in
figure 2 and then the Brauer--Weyl theory is not
working in the standard way for nilpotent representations.
\end{itemize}

\noindent
We believe that the correct possibility is the last one and we
shall present computational evidence for this.

We shall consider the next non--trivial case, $\epsilon^3=1$;
the Bratelli diagram for the centralizer is given in figure 5.
\begin{figure}
\vspace*{3cm}
\label{fig5}
\caption{}
\end{figure}
Let us compare for instance level 3 of figure 5 with
the decomposition $V^{\lambda} \otimes V^{\lambda} \otimes
V^{\lambda}$ depicted in figure 6.
\begin{figure}
\vspace*{7cm}
\label{fig6}
\caption{}
\end{figure}
The basis of $V^{\lambda}$ is $\{ e_i \}_{i=0}^2$ and
$M(e_{r_1} \otimes e_{r_2} \otimes e_{r_3}) = r_1+r_2+r_3$.
In the figure each dot stands for one of the linearly
independent states for each value of $M$. Dots linked by
vertical lines are connected by the action of the quantum group
generators on the space $V^{\lambda} \otimes V^{\lambda} \otimes
V^{\lambda}$, and so they have the same energy eigenvalue. They
also share the same eigenvalue of the quantum group generator
$\Delta^{(3)}(K)$, which is also given in the figure.
We realize that the different irreps appearing in figure 5
are one to one related with sets of irreps in figure 6
possessing the same value of $M$. In fact it can be explicitly
checked that the ``braiding" transformations $g_i$ defined by
(\ref{11}) close in the subspace defined by the same value of
$M$. From this we can conclude that if $C^{\lambda}_N(\epsilon)$
is generated by the $g_i$'s then Brauer--Weyl theory can not be
directly applied to the case of nilpotent irreps. Certainly this
result doesn't rule out the possibility of additional
generators; however the explicit analysis of the spectrum of the
hamiltonian $H(\lambda) + H^B(\lambda)$, presented in the first
part of this lecture, seems to indicate that this is not the
case.

For $\epsilon^3=1$ and a chain of 3 sites the dependences on
$\lambda$ of the energy eigenvalues $E_1,\ldots,E_9$ (see figure
6) of $H(\lambda) + H^B(\lambda)$ are given in figure 7.
\begin{figure}
\vspace*{6cm}
\label{fig7}
\caption{}
\end{figure}
By direct inspection of this figure we see that the energy
eigenvalues correponding to the same eigenvalue of $M$ have a
similar behaviour. It is worth mentioning that the Bratelli
diagrams in figures 2 and 5 can be derived
from a modification of the decomposition rule (\ref{16}). Indeed
if we supplement the irrep $\lambda$ with a new quantum number
$n \in {\bf N}$, and considering the fusion rule
\be
(\lambda_1,n_1) \otimes (\lambda_2,n_2) = \bigoplus_{r=0}^{N'-1}
(\lambda_1 \lambda_2 \epsilon^{-2r},n_1+n_2+r)
\label{19}
\ee

\noindent
we then obtain for $N'=2$ and 3 the Bratellis of figures
2 and 5 respectively.

This new quantum number $n$ is quite likely the Casimir of an
algebra whose representations are identical to the nilpotent
irreps of ${\cal U}_{\epsilon}(sl(2))$. This is indeed the case
of $N'=2$ and $\infty$, where this algebra is ${\cal U}_q(gl(1,1))$ \cite{SR}
and ${\cal U}(h_4)$ \cite{GSb} respectively.

Summarizing the content of this lecture:

\begin{enumerate}

\item We have obtained an integrable quantum group invariant
spin chain hamiltonian for nilpotent representations of \uq\ at
roots of unit.

\item We have defined the centralizer for nilpotent
representations $C^{\lambda}_N(\epsilon)$ and studied its
representation theory. It turns out that the irreps of the
centralizer generated by the $g_i$'s in equation (\ref{11}) are
one to one related (in the case $\epsilon^4=1$) to irreps of
$S_N$ characterized by ``corner" type Young diagrams.
\end{enumerate}

Many questions remain open. Among them it would be interesting
to provide a proof that $C^{\lambda}_N(\epsilon)$ is in fact
generated by the ``braiding" limit (\ref{10}) of the quantum
$R$--matrix $R^{\lambda \lambda}(u)$, and generalize to this
case the Brauer--Weyl theory. From a more speculative point of
view the situation concerning the centralizer we are facing here
strongly recalls the existence in CFT of extensions of chiral
algebras \cite{?}.

A more detalied presentation of the content of this lecture is
at present in preparation \cite{CGS}.

\end{document}